\documentclass[conference]{IEEEtran}

\usepackage{times}
\usepackage{mathptmx}
\usepackage{graphicx}
\usepackage{amsmath}
\begin{document}
%
\title{A Fully Distributed Opportunistic Network Coding Scheme for Cellular Relay Networks}
%
%
%

\author{\IEEEauthorblockN{Yulong Zou, Jia Zhu, and Baoyu Zheng}
\IEEEauthorblockA{Inst. of Signal Process. and Transm., Nanjing Univ. Post \& Telecomm., Nanjing, P. R. China\\
Email: \{Yulong.Zou, Jiazhu2010\}@gmail.com, zby@njupt.edu.cn}

}

\maketitle

\begin{abstract}
In this paper, we propose an opportunistic network coding (ONC) scheme in cellular relay networks, which operates depending on whether the relay decodes source messages successfully or not. A fully distributed method is presented to implement the proposed opportunistic network coding scheme without the need of any feedback between two network nodes. We consider the use of proposed ONC for cellular downlink transmissions and derive its closed-form outage probability expression considering cochannel interference in a Rayleigh fading environment. Numerical results show that the proposed ONC scheme outperforms the traditional non-cooperation in terms of outage probability. We also develop the diversity-multiplexing tradeoff (DMT) of proposed ONC and show that the ONC scheme obtains the \emph{full diversity} and an increased \emph{multiplexing gain} as compared with the conventional cooperation protocols.
\end{abstract}


\IEEEpeerreviewmaketitle

\section{Introduction}
%
%
%
%
\IEEEPARstart Next-generation cellular mobile networks, including International Mobile Telecommunications - Advanced and Beyond (IMT-Advanced and Beyond) [1], are expected to provide a peak download speed at 100Mbit/s (or higher) for mobile receptions and 1Gbit/s (or higher) for stationary receptions to meet continuously growing demand on mobile multimedia services (e.g., video-on-demand, mobile game/TV, and so on). Although the multiple-input multiple-output (MIMO) and orthogonal frequency division multiplexing (OFDM) are shown as effective methods to combat wireless fading and increase per-link throughput, they do not inherently mitigate the cochannel interference and fail to benefit cell-edge users significantly [2]. To that end, one promising solution is to use wireless relays which assist the transmissions between mobile users and a base station [3], [4].

Typically, relays are categorized into two broad types [2]: full-duplex relay and half-duplex relay, where the full-duplex relay refers to the relay capable of transmitting and receiving radio signals over the same channel, however, the half-duplex relay means that two channels (in terms of time/frequency) are required at a relay for transmitting and receiving signals. While full-duplex relays are attractive in terms of spectrum utilization, they are generally considered as impractical due to the significant difference in the power levels of incoming and outgoing signals. Thus, the half-duplex relay configuration is typically used, which, however, reduces the data rate since two channels are required to forward a message from source to destination [5]. To alleviate such an issue, two-way relays have been proposed by using a so-called physical network coding (PNC) that enables the transmission of two messages in two orthogonal channels [6]. However, the PNC requires complex symbol-level synchronization between distributed network nodes for signal combination and exact channel state information (CSI) of all network nodes at receiver side for signal decoding, which is challenging in practical systems.

In this paper, we investigate network coding without the complex symbol-level synchronization between distributed network nodes and propose an opportunistic network coding (ONC) scheme for cellular relay networks. It is worth mentioning that, although the proposed ONC is an opportunistic scheme that operates depending on whether the relay decodes source messages successfully or not, it can be implemented in a fully distributed manner without any feedback between two network nodes. This differs from the conventional network coding approaches where the cooperative users have to notify the destination whether or not they succeed in decoding each other's message so that the destination can perform the maximal ratio combining (MRC) among different network nodes in different situations of decoding outcomes (successful or not). In addition, differing from conventional network coding research (e.g., [7]-[9]), we study the ONC design and its diversity-multiplexing tradeoff (DMT) in cellular relay networks, where cochannel interference should be taken into account.

The rest of this paper is organized as follows. In Section II, we first present the system model of a cellular network with relay and then propose the ONC scheme for cellular downlink transmissions. Section III derives a closed-form outage probability expression of the proposed ONC scheme by considering cochannel interference in a Rayleigh fading environment, based on which a DMT analysis is also conducted. In Section IV, numerical results are presented to show the outage probability and DMT performance of the proposed ONC scheme and conventional cooperation protocols. Finally, Section VI provides concluding remarks.

\section{Proposed Opportunistic Network Coding}

\subsection{System Model}
\begin{figure}
  \centering
  {\includegraphics[scale=0.65]{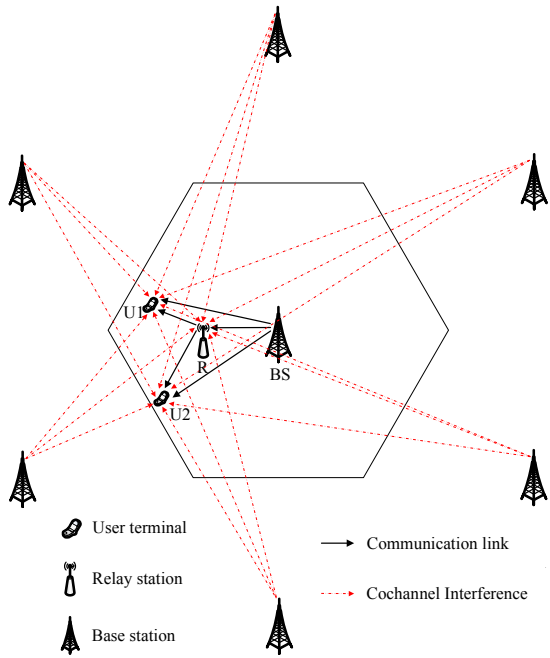}\\
  \caption{System model of a cellular relay network.}\label{Fig1}}
\end{figure}

Consider a cellular relay network as shown in Fig. 1, where a base station (BS) is located in the center of a cell and a relay station (RS) serves the BS. At present, such a cellular relay architecture has been adopted in the commercial wireless network IEEE 802.16j, where relay stations are allowed to communicate with BS and user terminals in one direction at a time (i.e., either uplink or downlink). In this paper, we assume that the channels are narrowband and modeled as Rayleigh fading, which correspond to ideal LTE OFDM subchannels.

Fig. 1 illustrates two cell-edge users (i.e., U1 and U2) that receive data from BS over downlinks with the assistance of one relay station (RS), where interferences (received at U1, U2 and RS) are from neighboring cochannel base stations. The reasons for considering the two-user cooperation are twofold. First, the two-user cooperation is simple for implementation in practical cellular systems, which is also shown as an effective means to improve wireless transmission performance [5]. Secondly, a general scenario with multiple users can be typically converted to the two-user cooperation by designing an additional grouping and partner selection protocol. In addition, this paper will focus on the cellular downlink transmission, while a similar analysis can be applied to the uplink.

\subsection{Proposed ONC Scheme}
Fig. 2 shows the proposed ONC encoding structure for cellular downlink transmissions, where BS intends to transmit $b_1 $ and $b_2$ to U1 and U2 in time slots $n$ and $n+1$, respectively. During time slot $n+2$, the information to be transmitted (from RS to U1 and U2) depends on whether RS succeeds in decoding $b_1 $ and $b_2 $ or not. Specifically, if RS decodes both $b_1 $ and $b_2 $ successfully, it transmits an XOR coded version of $b_1 $ and $b_2 $ to U1 and U2. If RS succeeds in decoding $b_1 $ (or $b_2 $) and fails to decode $b_2 $ (or $b_1 $), it transmits $b_1 $ (or $b_2 $) to U1 and U2. Otherwise, a null sequence is transmitted from RS in time slot $n+2$. Notice that, in practical systems, a cyclic redundancy check (CRC) code is generally used as forward error detection for every data package. Hence, RS can recognize whether or not it succeeds in decoding $b_1 $ or $b_2 $ by CRC checking, i.e., a successful CRC checking implies a correctly decoded outcome and vice versa.

We here consider that both $b_1$ and $b_2$ consist of CRC-encoded bits in two different CRC codes, as denoted by CRC\_1 and CRC\_2, respectively. Accordingly, in case that RS only forwards $b_1 $ or $b_2 $ during slot $n+2$ as shown in Fig. 2, both U1 and U2 are able to recognize through CRC checking that either $b_1 $ or $b_2 $ is transmitted, assuming a perfect error detection. In this way, the proposed ONC scheme can be implemented in a fully distributed manner without any feedback information between two nodes. The detailed illustration of this fully distributed implementation method will be presented in Section II-D. In addition, it is pointed out that, if there are multiple cell-edge users, we can pairwise them to generate multiple user pairs, where different user pairs proceed with the proposed opportunistic coding identically and independently of each other with different orthogonal channel groups. Since different user pairs can be differentiated through the identification of orthogonal channel groups, they can reuse the same set of CRC codes for the distributed implementation of the ONC scheme.
\begin{figure}
  \centering
  {\includegraphics[scale=0.8]{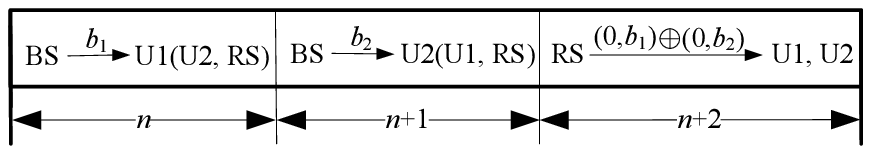}\\
  \caption{The encoding structure of proposed ONC for cellular downlink transmissions considering time division multiplexing (TDM).}\label{Fig2}}
\end{figure}

As shown in Fig. 2, BS first transmits $b_1$ to U1 in time slot $n$ and, at the same time, both U2 and RS overhear this transmission. Hence, the received signal at U1 in time slot $n$ can be expressed by
\begin{equation}\label{equa1}
y_1 (n) = \sqrt P h_{b1} (n)b_1  + \sum\limits_{k = 1}^K {\sqrt P g_{i_k 1} (n)I_{i_k } (n)}  + z_1 (n),
\end{equation}
where $P$ is transmit power, $h_{b1} (n)$ is the BS-U1 channel in time slot $n$, $K$ is the number of total cochannel interferers, $g_{i_k 1} (n)$ is the channel from $k$-th interferer to U1, $I_{i_k } (n)$ is the transmitted symbol of $k$-th interferer in time slot $n$, and $z_1 (n)$ is the additive white Gaussian noise (AWGN) at U1 with zero mean and variance $N_0$. Notice that subscripts $1$, $b$ and $i_k $ represent U1, BS and $k$-th interferer, respectively. We can similarly express the received signals at U2 and RS in time slot $n$ as
\begin{equation}\label{equa2}
y_2 (n) = \sqrt P h_{b2} (n)b_1  + \sum\limits_{k = 1}^K {\sqrt P g_{i_k 2} (n)I_{i_k } (n)}  + z_2 (n),
\end{equation}
and
\begin{equation}\label{equa3}
y_r (n) = \sqrt P h_{br} (n)b_1  + \sum\limits_{k = 1}^K {\sqrt P g_{i_k r} (n)I_{i_k } (n)}  + z_r (n),
\end{equation}
where $h_{b2} (n)$ and $h_{br} (n)$ are, respectively, the channels of BS-U2 and BS-R in time slot $n$, $g_{i_k 2} (n)$ and $g_{i_k r} (n)$ are the channels from $k$-th interferer to U2 and RS, respectively, and $z_2 (n)$ and $z_r (n)$ are AWGN with zero mean and variance $N_0$ at U2 and RS, respectively. Then, during next slot $n+1$, BS transmits $b_2 $ to U2 and, meanwhile, both U1 and RS overhear. Thus, the received signal at U2 in time slot $n+1$ is given by
\begin{equation}\label{equa4}
\begin{split}
y_2 (n + 1) =& \sqrt P h_{b2} (n + 1)b_2  + z_2 (n + 1)\\
&+ \sum\limits_{k = 1}^K {\sqrt P g_{i_k 2} (n + 1)I_{i_k } (n + 1)},
\end{split}
\end{equation}
where $h_{b2} (n + 1)$ is the BS-U2 channel in time slot $n+1$, $g_{i_k 2} (n + 1)$ is the channel from $k$-th interferer to U2, and $z_2 (n + 1)$ is the AWGN at U2 with zero mean and variance $N_0$. Meanwhile, the received signals at U1 and RS are, respectively, given by
\begin{equation}\label{equa5}
\begin{split}
y_1 (n + 1) =& \sqrt P h_{b1} (n + 1)b_2  + z_1 (n + 1)\\
& + \sum\limits_{k = 1}^K {\sqrt P g_{i_k 1} (n + 1)I_{i_k } (k + 1)} ,
\end{split}
\end{equation}
and
\begin{equation}\label{equa6}
\begin{split}
y_r (n + 1) =& \sqrt P h_{br} (n + 1)b_2  + z_r (n + 1) \\
& + \sum\limits_{k = 1}^K {\sqrt P g_{i_k r} (n + 1)I_{i_k } (n + 1)}.
\end{split}
\end{equation}
where $h_{b1} (n+1)$ and $h_{br} (n+1)$ are, respectively, the channels of BS-U1 and BS-R in time slot $n+1$, $g_{i_k 1} (n+1)$ and $g_{i_k r} (n+1)$ are the channels from $k$-th interferer to U2 and RS, respectively, and $z_1 (n+1)$ and $z_r (n+1)$ are AWGN at U2 and RS, respectively.

\subsection{Decoding Structure at a Relay Station}
\begin{figure}
  \centering
  {\includegraphics[scale=0.8]{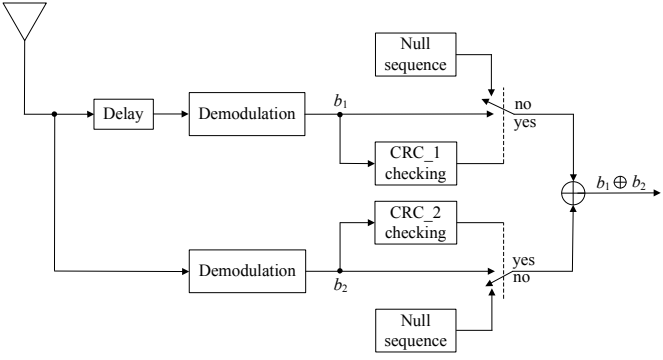}\\
  \caption{A receiver structure at the relay station.}\label{Fig3}}
\end{figure}

In what follows, we present a decoder design of the proposed ONC at the relay station. As shown in Fig. 3, if CRC\_1 and CRC\_2 checking both pass, $b_1  \oplus b_2 $ will be transmitted at RS in time slot $n+2$. If RS fails in both CRC\_1 and CRC\_2 checking, a null sequence will be transmitted. One can see from Fig. 3 that, if RS succeeds in CRC\_1 checking (or, CRC\_2 checking) and fails to pass CRC\_2 checking (or, CRC\_1 checking), it transmits $b_1 $ (or, $b_2$) in time slot $n+2$. For notational convenience, let $\theta  = 1$, $2$, $3$, and $4$, respectively, denote the above-mentioned four cases, i.e., CRC\_1 and CRC\_2 checking both pass, CRC\_1 checking passes and CRC\_2 checking fails, CRC\_1 checking fails and CRC\_2 checking passes, and both CRC\_1 and CRC\_2 checking fail. It is assumed that a failed CRC checking indicates that an outage event occurs. Hence, given a data transmission rate $R$ over downlink, we can describe events $\theta  = 1$, $2$, $3$, and $4$ as (in an information-theoretic sense [5])
\begin{equation}\label{equa7}
\begin{split}
  & \theta  = 1:\quad I_{br} (n) > R{\textrm{ and }}I_{br} (n + 1) > R \\
  & \theta  = 2:\quad I_{br} (n) > R{\textrm{ and }}I_{br} (n + 1) < R \\
  & \theta  = 3:\quad I_{br} (n) < R{\textrm{ and }}I_{br} (n + 1) > R  \\
  & \theta  = 4:\quad I_{br} (n) < R{\textrm{ and }}I_{br} (n + 1) < R, \\
\end{split}
\end{equation}
where $I_{br} (n)$ and $I_{br} (n + 1)$ are the mutual information from BS to RS in time slots $n$ and $n+1$, respectively. Following Eq. (3) and considering coherent detection, $I_{br} (n)$ are given by
\begin{equation}\label{equa8}
I_{br} (n) = \frac{2}
{3}\log _2 (1 + \frac{{|h_{br} (n)|^2 \gamma }}
{{\sum\limits_{k = 1}^K {|g_{i_k r} (n)|^2 \gamma }  + 1}}),
\end{equation}
where $\gamma  = P/N_0 $ is signal-to-noise ratio (SNR) and factor $\frac{2}{3}$ in front of $\log _2 ( \cdot )$ is due to the fact that three time slots are used for transmitting two information symbols $b_1 $ and $b_2 $. Similarly, from Eq. (6), we can obtain $I_{br} (n + 1)$ with coherent detection as
\begin{equation}\label{equa9}
I_{br} (n + 1) = \frac{2}{3}\log _2 (1 + \frac{{|h_{br} (n + 1)|^2 \gamma }}{{\sum\limits_{k = 1}^K {|g_{i_k r} (n + 1)|^2 \gamma }  + 1}}).
\end{equation}
Meanwhile, given $\theta  = 1$, $2$, $3$, and $4$, the received signal at U1 in time slot $n+2$ is given by Eq. (10) at the top of the following page,
\begin{figure*}
\begin{equation}\label{equa10}
y_1 (n + 2) = \begin{cases}
  \sqrt P h_{r1} (n + 2)(b_1  \oplus b_2 ) + \sum\limits_{k = 1}^K {\sqrt P g_{i_k 1} (n + 2)I_{i_k } (k + 2)}  + z_1 (n + 2),&\theta  = 1\\
  \sqrt P h_{r1} (n + 2)b_1  + \sum\limits_{k = 1}^K {\sqrt P g_{i_k 1} (n + 2)I_{i_k } (k + 2)}  + z_1 (n + 2),&\theta  = 2\\
  \sqrt P h_{r1} (n + 2)b_2  + \sum\limits_{k = 1}^K {\sqrt P g_{i_k 1} (n + 2)I_{i_k } (k + 2)}  + z_1 (n + 2),&\theta  = 3\\
  \sqrt P h_{r1} (n + 2)b + \sum\limits_{k = 1}^K {\sqrt P g_{i_k 1} (n + 2)I_{i_k } (k + 2)}  + z_1 (n + 2),&\theta  = 4 \end{cases}
\end{equation}
\end{figure*}
where $h_{r1} (n + 2)$ is the R-U1 channel in time slot $n+2$, $g_{i_k 1} (n + 2)$ is the channel from $k$-th interferer to U2, $z_1 (n + 2)$ is the AWGN at U2 with zero mean and variance $N_0$, and $b$ represents a null sequence transmitted at RS.

\subsection{Decoding Structure at User Terminals}
\begin{figure}
  \centering
  {\includegraphics[scale=0.72]{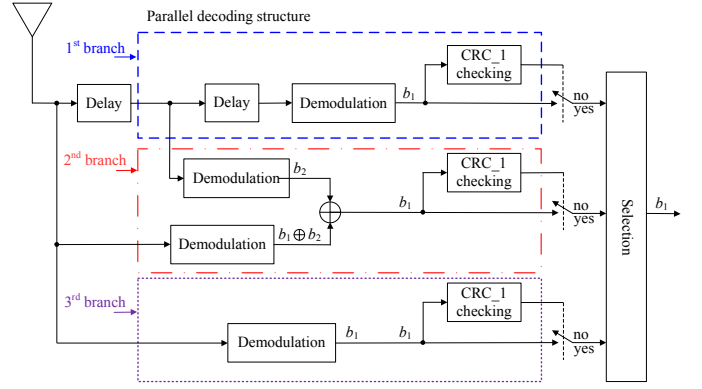}\\
  \caption{A parallel decoder structure at U1 in decoding $b_1 $. The process at U2 in decoding $b_2$ is similar.}\label{Fig4}}
\end{figure}

In this subsection, we present the decoding process of proposed ONC scheme at user terminals. We focus on the details of the decoder structure at U1 in decoding $b_1 $, and a similar design can be applied to U2 in decoding $b_2 $. As shown in Fig. 4, we utilize three parallel branches at U1 in decoding $b_1 $, where the first branch is to decode the direct transmission of $b_1 $ from BS to U1, the second branch is to combine the transmissions from BS (i.e., $b_2 $) and RS (e.g., $b_1  \oplus b_2 $) to U1, and the third branch is used to demodulate the possible transmission of $b_1 $ from RS. Typically, the branch that passes CRC\_1 checking is selected as the decoder output at U1. Moreover, if more than one branch succeed in CRC checking, we can choose one of the successful branches as the output. One can observe from Fig. 4 that, in the proposed ONC scheme, signal transmissions from different network nodes at different slots are demodulated separately at receiver without the signal combination between different transmissions, which can avoid the complex symbol-level synchronization issue and shows the advantage of the proposed opportunistic network coding over conventional PNC [6]. Also, Fig. 4 shows that U1 can decode $b_1$ locally without any feedback information from RS, implying that the proposed ONC scheme is implemented in a fully distributed manner. As shown in Fig. 4, given $\theta  = 1$ (i.e., RS decodes both $b_1 $ and $b_2$), U1 would possibly recover $b_1 $ either from the first branch or second branch. Thus, in this case, the conditional mutual information from BS to U1 is given by
\begin{equation}\label{equa11}
I_{b1} (\theta  = 1) = \max \{ I_{b1} (n),\min [I_{b1} {\textrm{(}}n + 1{\textrm{)}},I_{r1} (n + 2)]\},
\end{equation}
where $I_{b1} (n)$, $I_{b1} {\textrm{(}}n + 1{\textrm{)}}$, and $I_{r1} (n + 2)$ are the mutual information from BS to U1 in time slot $n$, from BS to U1 in time slot $n+1$, and from RS to U1 in time slot $n+2$, respectively. Following Eqs. (1) and (5), we, respectively, obtain the mutual information $I_{b1} (n)$ and $I_{b1} {\textrm{(}}n + 1{\textrm{)}}$ as
\begin{equation}\label{equa12}
I_{b1} (n) = \frac{2}{3}\log _2 (1 + \frac{{|h_{b1} (n)|^2 \gamma }}{{\sum\limits_{k = 1}^K {|g_{i_k 1} (n)|^2 \gamma }  + 1}}),
\end{equation}
and
\begin{equation}\label{equa13}
I_{b1} (n + 1) = \frac{2}{3}\log _2 (1 + \frac{{|h_{b1} (n + 1)|^2 \gamma }}{{\sum\limits_{k = 1}^K {|g_{i_k 1} (n + 1)|^2 \gamma }  + 1}}).
\end{equation}
Similarly, one can easily obtain the mutual information $I_{r1} (n + 2)$ from Eq. (10) as
\begin{equation}\label{equa14}
I_{r1} (n + 2) = \frac{2}{3}\log _2 (1 + \frac{{|h_{r1} (n + 2)|^2 \gamma }}{{\sum\limits_{k = 1}^K {|g_{i_k 1} (n + 2)|^2 } \gamma  + 1}}).
\end{equation}
Given $\theta  = 2$ (i.e., RS decodes $b_1 $, but fails to decode $b_2$), U1 would possibly succeeds in decoding $b_1 $ either from the first branch or third branch as shown in Fig. 4. Hence, in given case $\theta  = 2$, the conditional mutual information from BS to U1 is obtained as
\begin{equation}\label{equa15}
I_{b1} (\theta  = 2) = \max \{ I_{b1} (n),I_{r1} (n + 2)\}.
\end{equation}
Finally, either event $\theta  = 3$ or $\theta  = 4$ occurs, U1 can rely on the first branch only to decode $b_1 $. Thus, given case $\theta  = 3$ or $\theta  = 4$, the corresponding conditional information from BS to U1 is given by
\begin{equation}\label{equa16}
I_{b1} (\theta  = 3) = I_{b1} (\theta  = 4) = I_{b1} (n),
\end{equation}
where $I_{b1} (n)$ is given by Eq. (12). Now, we complete the signal modeling for the decoding process of proposed ONC scheme.

\section{Performance Analysis of Proposed ONC Scheme over Rayleigh Fading Channels}
In this section, we focus on the performance analysis of the transmission from BS to U1 and the BS-U2 transmission has similar performance results. We first examine outage probability of the ONC scheme, followed by a DMT analysis. As is known [5], an outage event occurs when the channel capacity falls below a predefined data rate $R$. Hence, an outage probability of the ONC scheme is given by
\begin{equation}\label{equa17}
\begin{split}
&{\textrm{Pout}}_{{\textrm{ONC}}}  = \Pr [I_{b1}  < R] \\
&= \sum\limits_{k = 1,2,3,4} {\Pr (\theta  = k)\Pr [I_{b1} (\theta  = k) < R]}.
\end{split}
\end{equation}
Using Eqs. (7) - (9), we can obtain term $\Pr (\theta  = 1)$ as
\begin{equation}\label{equa18}
\begin{split}
\Pr (\theta  = 1) &= \Pr [I_{br} (n) > R]\Pr [I_{br} (n + 1) > R] \\
& \doteq \Pr [\frac{{|h_{br} (n)|^2 }}
{{\sum\limits_{k = 1}^K {|g_{i_k r} (n)|^2 } }} > 2^{3R/2}  - 1]\\
&\quad\times\Pr [\frac{{|h_{br} (n + 1)|^2 }}
{{\sum\limits_{k = 1}^K {|g_{i_k r} (n + 1)|^2 } }} > 2^{3R/2}  - 1], \\
\end{split}
\end{equation}
where the second equation is obtained by ignoring the noise. This is valid when the interference becomes a dominant concern, e.g., in interference-limited systems. Note that random variables $|h_{br} (n)|^2 $, $|h_{br} (n + 1)|^2 $, $|g_{i_k r} (n)|^2 $, and $|g_{i_k r} (n + 1)|^2 $ follow exponential distributions and are independent of each other. Thus, we can further obtain $\Pr (\theta  = 1)$ as Eq. (19) at the top of this page,
\begin{figure*}
\begin{equation}\label{equa19}
\Pr (\theta  = 1) = \begin{cases}
  \left(\sum\limits_{k = 1}^K {\dfrac{{\lambda _{br{\textrm{-}}i_k r} }}
{{\lambda _{br{\textrm{-}}i_k r}  + (2^{3R/2}  - 1)}}\prod\limits_{j = 1,j \ne k}^K {\dfrac{{\lambda _{br{\textrm{-}}i_k r}^{ - 1} }}{{\lambda _{br{\textrm{-}}i_k r}^{ - 1}  - \lambda _{br{\textrm{-}}i_j r}^{ - 1} }}} } \right)^{2} ,&\lambda _{br{\textrm{-}}i_1 r}  \ne  \cdots  \ne \lambda _{br{\textrm{-}}i_K r} \\
\\
  \left(\dfrac{{\lambda _{br{\textrm{-}}i_k r} }}{{\lambda _{br{\textrm{-}}i_k r}  + (2^{3R/2}  - 1)}}\right)^{2K} ,&\lambda _{br{\textrm{-}}i_1 r}  = \cdots  = \lambda _{br{\textrm{-}}i_K r} \\
 \end{cases}
\end{equation}
\end{figure*}
where $\sigma _{br}^2  = E(|h_{br} (n)|^2 ) = E(|h_{br} (n + 1)|^2 )$, $\sigma _{i_k r}^2  = E(|h_{i_k r} (n)|^2 ) = E(|h_{i_k r} (n + 1)|^2 )$, and $\lambda _{br{\textrm{-}}i_k r}  = \sigma _{br}^2 /\sigma _{i_k r}^2 $ is viewed as the signal-to-interference ratio (SIR) of the channel gain from BS to RS to that from $k$-th interferer to RS. We can similarly determine terms $\Pr (\theta  = 2)$, $\Pr (\theta  = 3)$, and $\Pr (\theta  = 4)$ in closed-form. Besides, following Eqs. (11) - (14) and ignoring noise, we obtain term $\Pr [I_{b1} (\theta  = 1) < R]$ in Eq. (17) as
\begin{equation}\label{equa20}
\begin{split}
&\Pr [I_{b1} (\theta  = 1) < R]  = \Pr [\frac{{|h_{b1} (n)|^2 }}{{\sum\limits_{k = 1}^K {|g_{i_k 1} (n)|^2 } }} < 2^{3R/2}  - 1]\\
&\quad\quad\quad \quad\quad-  \Pr [\frac{{|h_{b1} (n)|^2 }}{{\sum\limits_{k = 1}^K {|g_{i_k 1} (n)|^2 } }} < 2^{3R/2}  - 1] \\
&\quad\quad\quad\quad\quad\quad\times\Pr [\frac{{|h_{b1} (n + 1)|^2 }}{{\sum\limits_{k = 1}^K {|g_{i_k 1} (n + 1)|^2 } }} > 2^{3R/2}  - 1]\\
&\quad\quad\quad\quad\quad\quad\times \Pr [\frac{{|h_{r1} (n + 2)|^2 }}{{\sum\limits_{k = 1}^K {|g_{i_k 1} (n + 2)|^2 } }} > 2^{3R/2}  - 1],
\end{split}
\end{equation}
where $\Pr [\frac{{|h_{b1} (n)|^2 }}{{\sum\limits_{k = 1}^K {|g_{i_k 1} (n)|^2 } }} < 2^{3R/2}  - 1]$, $\Pr [\frac{{|h_{b1} (n + 1)|^2 }}{{\sum\limits_{k = 1}^K {|g_{i_k 1} (n + 1)|^2 } }} > 2^{3R/2}  - 1]$, and $\Pr [\frac{{|h_{r1} (n + 2)|^2 }}{{\sum\limits_{k = 1}^K {|g_{i_k 1} (n + 2)|^2 } }} > 2^{3R/2}  - 1]$ can be easily determined in closed-form. Similarly to Eq. (20), we can also obtain closed-form solutions to $\Pr [I_{b1} (\theta  = 2) < R]$, $\Pr [I_{b1} (\theta  = 3) < R]$ and $\Pr [I_{b1} (\theta  = 4) < R]$. So far, we have completed the closed-form outage probability analysis for proposed ONC scheme, based on which the DMT will be developed in the following. Note that the traditional diversity gain is defined as $d =  - \mathop {\lim }\limits_{{\textrm{SNR}} \to \infty } \frac{{\log P_e ({\textrm{SNR}})}}{{\log {\textrm{SNR}}}}$ [12] where ${\textrm{SNR}}$ is SNR and $P_e $ represents bit error rate, which is not applicable here since the interference, rather than AWGN noise, becomes a dominant concern in determining the transmission performance, as shown in Eqs. (18) and (20). Therefore, we present a generalized diversity gain as an asymptotic ratio of the outage performance to SIR $\lambda _{b1{\textrm{-}}i_1 1}  = \sigma _{b1}^2 /\sigma _{i_1 1}^2 $ with $\lambda _{b1{\textrm{-}}i_1 1}  \to \infty $, where $\sigma _{i_1 1}^2  = E(|h_{i_1 1} |^2 )$ is the average gain of the channel from the $1^{\textrm{st}}$ interferer to U1. Accordingly, the diversity gain of proposed ONC scheme is given by
\begin{equation}\label{equa21}
d_{{\textrm{ONC}}}  =  - \mathop {\lim }\limits_{\lambda _{b1{\textrm{-}}i_1 1}  \to \infty } \frac{{\log ({\text{Pout}}_{{\textrm{ONC}}} )}}{{\log (\lambda _{b1{\textrm{-}}i_1 1} )}}.
\end{equation}
Meanwhile, the multiplexing gain $r$ is defined as
\begin{equation}\label{equa22}
r = \mathop {\lim }\limits_{\lambda _{b1{\textrm{-}}i_1 1}  \to \infty }\frac{R(\lambda _{b1{\textrm{-}}i_1 1})}{\log (\lambda _{b1{\textrm{-}}i_1 1} )}.
\end{equation}
Following the generalized DMT definition as given by Eqs. (21) and (22), we can obtain the DMT performance of proposed ONC scheme as
\begin{equation}\label{equa23}
d_{{\textrm{ONC}}}  + 3r = 2.
\end{equation}
One can observe from Eq. (23) that a diversity gain $d_{{\textrm{ONC}}}  = 2$ is achieved as $r \to 0$ and, on the other hand, a maximum multiplexing gain of two-third (i.e., $ r = 2/3$) can be achieved as the diversity gain approaches to zero.

\section{Numerical Results}
\begin{figure}
  \centering
  {\includegraphics[scale=0.6]{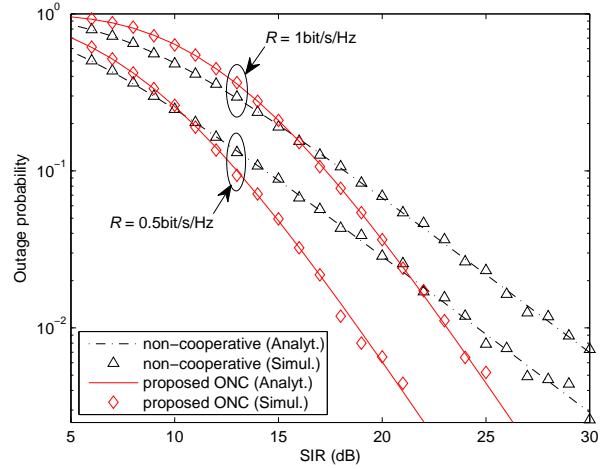}\\
  \caption{Outage probability versus signal-to-interference ratio (SIR) of the non-cooperation and proposed ONC schemes with $K = 7$ and $\lambda _{b1{\textrm{-}}i_k 1}  = \lambda _{br{\textrm{-}} i_k r}  = \lambda _{r1{\textrm{-}}i_k 1}  = \lambda _{rb{\textrm{-}}i_k b}  = \lambda _{1r{\textrm{-}}i_k r}  = \lambda _{1b{\textrm{-}}i_k b}  = \lambda _{21{\textrm{-}}i_k 1}  = \lambda _{2r{\textrm{-}}i_k r} $.}\label{Fig5}}
\end{figure}

Fig. 5 shows the outage probability versus the signal-to-interference radio (SIR) of the non-cooperation and proposed ONC schemes for different data rates, where the simulation results are also given. It is shown from Fig. 5 that the simulation results match analytical results very well. One can observe from Fig. 5 that in low SIR regions, the proposed ONC scheme performs worse than the non-cooperation in terms of outage probability for both $R=0.5 {\textrm{bit/s/Hz}}$ and $R=1{\textrm{bit/s/Hz}}$. This is because that a half-duplex relay constraint is considered for the ONC scheme, which sacrifices the spectrum efficiency (also known as multiplexing gain) to achieve the diversity gain. However, in higher SIR regions, the benefits achieved from diversity gain overtake costs due to the half-duplex relay constraint and the outage probability performance of the ONC scheme becomes better than that of the non-cooperation.
\begin{figure}
  \centering
  {\includegraphics[scale=0.6]{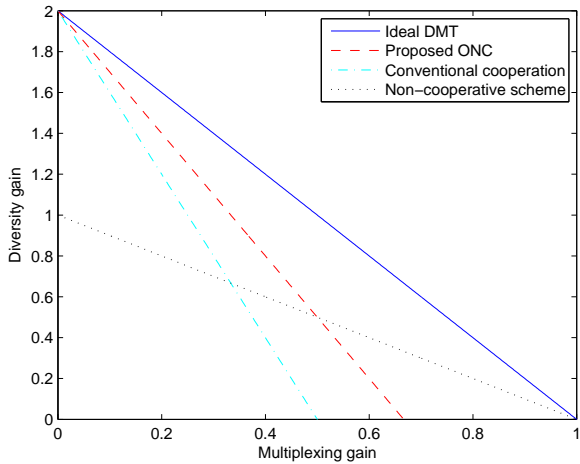}\\
  \caption{Diversity-multiplexing tradeoffs of the non-cooperation, conventional cooperation and proposed ONC schemes.}\label{Fig6}}
\end{figure}

Fig. 6 compares the DMT performance of the non-cooperation, conventional cooperation protocols, and proposed ONC scheme. As shown in Fig. 6, as the multiplexing gain approaches zero, the non-cooperation achieves a diversity gain of only one; however the conventional cooperation and proposed ONC schemes obtain the full diversity gain of two, showing the advantage of cooperation over non-cooperation. On the other hand, one can also see from Fig. 6 that as the diversity gain decreases to zero, the conventional cooperation schemes achieve a maximum multiplexing gain of one-half. In contrast, the proposed ONC scheme obtains a maximum multiplexing gain of two-third, which is better than the conventional cooperation.

\section{Conclusion}
In this paper, we investigated opportunistic network coding for cellular relay networks and proposed a fully distributed ONC scheme. We derived a closed-form outage probability expression of the proposed ONC scheme over Rayleigh fading channels. Numerical outage probability results showed that the ONC scheme performs better than the non-cooperation. We also studied the DMT performance of proposed ONC scheme and showed that the proposed ONC strictly outperforms the conventional cooperation in terms of DMT performance.

\ifCLASSOPTIONcaptionsoff
  \newpage
\fi

\end{document}